\documentclass[12pt]{amsart}
\usepackage{amssymb}
\usepackage{isolatin1}    
\newtheorem{thm}{Theorem}[section]
\newtheorem{lem}[thm]{Lemma}
\newtheorem{cor}[thm]{Corollary}
\newtheorem{prop}[thm]{Proposition}
\newtheorem{ex}[thm]{Example}

\newtheorem*{prob*}{Open problem}

\theoremstyle{definition}

\newtheorem{defi}[thm]{Definition}

\theoremstyle{remark}

\newtheorem{rem}[thm]{Remark}
\newtheorem*{rem*}{Remark}

\DeclareMathOperator{\id}{id}
\DeclareMathOperator{\s}{span}

\newcommand{\kringel}{\mathbin{\raise1pt\hbox{$\scriptstyle\circ$}}} 
\newcommand{\pkt}{\mathbin{\raise0pt\hbox{$\scriptstyle\bullet$}}}

\newcommand{\C}{\mathbb{C}}

\newcommand{\Z}{\mathbb{Z}}

\newcommand{\ad}{\mathop{\rm ad}}
\newcommand{\Ad}{\mathop{\rm Ad}}
\newcommand{\End}{\mathop{\rm End}}
\newcommand{\Der}{\mathop{\rm Der}}

\newcommand{\La}{\mathfrak{a}}
\newcommand{\Lb}{\mathfrak{b}}

\newcommand{\Lf}{\mathfrak{f}}
\newcommand{\Lg}{\mathfrak{g}}
\newcommand{\Lh}{\mathfrak{h}}

\newcommand{\Ll}{\mathfrak{l}}
\newcommand{\Ln}{\mathfrak{n}}

\newcommand{\Lr}{\mathfrak{r}}
\newcommand{\Ls}{\mathfrak{s}}

\newcommand{\al}{\alpha}
\newcommand{\be}{\beta}
\newcommand{\ga}{\gamma}

\newcommand{\la}{\lambda}
\newcommand{\om}{\omega}

\newcommand{\Om}{\Omega}

\newcommand{\ra}{\rightarrow}

\renewcommand{\phi}{\varphi}

\begin{document}


\title[Novikov structures]{Novikov structures on solvable Lie algebras}

\author[D. Burde]{Dietrich Burde}
\author[K. Dekimpe]{Karel Dekimpe}
\address{Fakult\"at f\"ur Mathematik\\
Universit\"at Wien\\
  Nordbergstr. 15\\
  1090 Wien \\
  Austria} 
\email{dietrich.burde@univie.ac.at}
\address{Katholieke Universiteit Leuven\\
Campus Kortrijk\\
8500 Kortrijk\\
Belgium}
\date{\today}
\email{Karel.Dekimpe@kulak.ac.be}

\thanks{The second author thanks the Erwin Schr\"odinger Institute for its hospitality and support}

\begin{abstract}
We study Novikov algebras and Novikov structures on finite-dimensional Lie algebras.
We show that a Lie algebra admitting a Novikov structure must be solvable. Conversely
we present an example of a nilpotent $2$-step solvable Lie algebra without any Novikov structure.
We construct Novikov structures on certain Lie algebras via classical $r$-matrices
and via extensions. In the latter case we lift Novikov structures on an abelian Lie
algebra $\La$ and a Lie algebra $\Lb$ to certain extensions of $\Lb$ by $\La$.
We apply this to prove the existence of affine and Novikov structures on several
classes of $2$-step solvable Lie algebras. In particular we generalize a well known result of
Scheuneman concerning affine structures on $3$-step nilpotent Lie algebras.
\end{abstract}

\maketitle

\section{Introduction}

Novikov algebras arise in several contexts in mathematics and physics.
They were firstly introduced in the study of
Hamiltonian operators concerning integrability of certain nonlinear partial
differential equations \cite{GD}. They also appear in connection with Poisson
brackets of hydrodynamic type and operator Yang-Baxter equations \cite{BN}.
Furthermore, if $A$ is a Novikov algebra, then
$\widehat{A}=A \otimes \C[t,t^{-1}]$ is equipped with a Lie bracket via
\[
[x\otimes t^m, y\otimes t^n]=mxy \otimes t^{m+n-1}-n yx \otimes t^{m+n-1}
\]
for $x,y \in A$ and $m,n\in \Z$.
The Lie algebra $\widehat{A}$ induces a class of vertex algebras.
This is of interest in conformal field theory, see \cite{FHL}. \par
A Novikov algebra is a special case of an LSA - a left-symmetric algebra.
LSAs arise in the study of affine manifolds, affine crystallographic groups,
convex homogeneous cones, and the areas mentioned above.
An LSA is a Lie-admissible algebra: the commutator defines
a Lie algebra. One may ask which Lie algebras can arise that way.
This is the existence question of affine structures on Lie algebras.
It is quite difficult and has been studied a lot, see \cite{BU1}.
For Novikov structures the existence question is more accessible. 
We are able to prove interesting results. On the other hand we study 
natural construction methods of left-symmetric and Novikov algebras.
For example, it is known in physics that a classical $r$-matrix on a Lie algebra
$\Lg$ induces a left-symmetric product on the vector space of $\Lg$.
We modify this idea to obtain also Novikov structures. 
Another natural construction of affine structures is the lifting of given
affine structures on an abelian Lie algebra $\La$ and a Lie algebra
$\Lb$ to an affine structure on the extension of $\Lb$ by $\La$.
We describe the conditions for lifting affine and Novikov structures.\\
The paper is organized as follows: in the first section we introduce affine and 
Novikov structures and show that a Lie algebra admitting a Novikov structure must be solvable.
Then we give an example of an $8$-dimensional $4$-step nilpotent and
$2$-step solvable Lie algebra which does not admit any Novikov structure. \\
In the second section we show how to construct Novikov structures on 
certain Lie algebras via classical $r$-matrices. In low dimensions
all solvable Lie algebras admit a Novikov structure that way. \\
In the third section we determine the conditions for lifting affine
and Novikov structures via extensions. We apply this to give a new
proof of Scheunemann's result concerning affine structures on
$3$-step nilpotent Lie algebras. For Novikov structures we obtain
a similar result, if the Lie algebra is $2$- or $3$-generated.
However, the proof does not work in general for Novikov structures. 
But for a certain class of $2$-step solvable Lie algebras we are able to 
obtain Novikov structures via extensions. Among them is the class of
filiform Lie algebras with abelian commutator algebra. \\
In the last section we show by cohomological considerations
that the lifting procedure via extensions can be essentially reduced to nilpotent 
Lie algebras. As an application we can prove a generalization of 
Scheuneman's result to certain $2$-step solvable Lie algebras.
 
\section{Affine and Novikov structures}

Let $k$ be a field of characteristic zero. A Novikov algebra and, more
generally, an LSA is defined as follows:

\begin{defi}
An algebra $(A,\cdot)$ over $k$ with product $(x,y) \mapsto x\cdot y$
is called {\it left-symmetric algebra (LSA)}, if the product is
left-symmetric, i.e., if the identity
\begin{equation}\label{lsa1}
x\cdot (y\cdot z)-(x\cdot y)\cdot z= y\cdot (x\cdot z)-(y\cdot x)\cdot z
\end{equation}
is satisfied for all $x,y,z \in A$. The algebra is called {\it Novikov}, if
in addition
\begin{equation}
(x\cdot y)\cdot z=(x\cdot z)\cdot y
\end{equation}
is satisfied.
\end{defi}

Denote by $L(x), R(x)$ the right respectively left multiplication operator in
the algebra  $(A,\cdot)$. Then a LSA is a Novikov algebra if the right
multiplications commute:
\begin{align*}
[R(x),R(y)] & = 0
\end{align*}

It is well known that LSAs are Lie-admissible algebras: the commutator
defines a Lie bracket. The associated Lie algebra then is said to admit
a left-symmetric structure, or affine structure.

\begin{defi}\label{affine}
An {\it affine structure} on a Lie algebra
$\Lg$ over $k$ is a left-symmetric product $\Lg \times \Lg \rightarrow \Lg$
satisfying 
\begin{equation}\label{lsa2}
[x,y]=x\cdot y -y\cdot x
\end{equation}
for all $x,y,z \in \Lg$. 
If the product is Novikov, we say that $\Lg$ admits a {\it Novikov structure}.
\end{defi}     
 
An affine structure on a Lie algebra $\Lg$ corresponds to a left-invariant
affine structure on a connected, simply connected Lie group $G$ with Lie algebra
$\Lg$. Such structures play an important role for affine crystallographic groups
and affine manifolds, see \cite{BU1}.
It is well known that not every Lie algebra admits an affine structure. 
A finite-dimensional Lie algebra $\Lg$ over $k$ with $[\Lg,\Lg]=\Lg$
does not admit an affine structure. But also not every solvable Lie algebra
admits an affine structure. There exist complicated examples of even nilpotent Lie algebras 
which do not admit an affine structure, see \cite{BEN}, \cite{BU1}. 
These examples are of dimension $10,11,12,13$.
On the other hand there are many classes of solvable and nilpotent Lie algebras which do
admit an affine structure: every positively graded Lie algebra admits an affine structure, 
and every $2$ and $3$-step nilpotent Lie algebra admits an affine structure.\\
We need some notations for solvable and nilpotent Lie algebras.
Denote the terms of the commutator series
by $\Lg^{(1)}=\Lg$, $\Lg^{(i+1)}=[\Lg^{(i)},\Lg^{(i)}]$. 
Then $\Lg$ is called $p$-step solvable if $\Lg^{(p+1)}=0$.
Denote the terms of the lower central series by 
$\Lg^1=\Lg$, $\Lg^{i+1}=[\Lg,\Lg^{i}]$. The Lie algebra $\Lg$
is called $p$-step nilpotent, if $\Lg^{p+1}=0$.

\begin{rem}
A three-step nilpotent Lie algebra is two-step solvable: $\Lg^4=0$ and the Jacobi identity
imply $\Lg^{(3)}=0$, since
$$
[[x,y],[z,w]]= [w,[z,[x,y]] - [z,[w,[x,y]]] = 0
$$  
\end{rem}

The following two lemmas are elementary but useful. We leave the proof to the
reader.

\begin{lem}\label{nov}
Let $(A,\cdot)$ be an algebra product on the vector space of a Lie algebra
$\Lg$. Then $(A,\cdot)$ defines a Novikov structure on $\Lg$ if and only if
\begin{align*}
 L(x)-R(x) & = \ad (x) \\
[L(x),L(y)] & = L([x,y])\\
[R(x),R(y)] & = 0
\end{align*}
for all $x,y \in \Lg$. 
\end{lem}

\begin{lem}
Let $(A,\cdot)$ be a Novikov algebra. Then the following two identities hold for
all $x,y,z \in A$:
\begin{align*}
[x,y]\cdot z + [y,z]\cdot x + [z,x]\cdot y & = 0 \\
x \cdot [y,z] + y\cdot [z,x] + z\cdot [x,y] & = 0 \\
\end{align*}
\end{lem}

\begin{defi}
An affine structure on a Lie algebra $\Lg$ is called {\it complete}, if the
right multiplications $R(x)$ are nilpotent for all $x\in \Lg$.
\end{defi}  

The existence question of affine structures is very difficult in general.
For Novikov algebras this question is more accessible.
First, it follows from a result in \cite{ZEL} 
that the underlying Lie algebra of a Novikov algebra is solvable:

\begin{prop}
Any finite-dimensional Lie algebra admitting a Novikov structure is solvable.
\end{prop}

\begin{proof}
Assume that $\Lg$ admits a Novikov structure given by
$(x,y)\mapsto x\cdot y$. Denote by $R(x)$ the right multiplication 
in the Novikov algebra $A$, i.e., $R(x)(y)=y\cdot x$. The algebra $A$
is called right-nilpotent if $R_A=\{R(x) \mid x\in A \}$ satisfies
$R_A^n=0$ for some $n\ge 1$. Let $I,J$ be two right-nilpotent ideals
in $A$. Since $[R(x),R(y)]=0$ the sum $I+J$ is also a right-nilpotent
ideal. Since $A$ is finite-dimensional there exists a largest right-nilpotent
ideal of $A$, denoted by $N(A)$. Now $N(A)$ is a complete left-symmetric
algebra since its right multiplications are nilpotent.
It is known that the Lie algebra of a complete LSA is solvable, see \cite{SEG}.
Hence the Lie algebra $\Lh$ of $N(A)$ is solvable. 
On the other hand $A/N(A)$ is a direct sum of fields.
This was proved in \cite{ZEL}. It follows that the Lie algebra of $A/N(A)$ is abelian.
Hence $\Lg/\Lh$ is abelian, and $\Lh$ is solvable. It follows that $\Lg$ is solvable.
\end{proof}

\begin{prop}
Let $\Lg$ be a finite-dimensional two-step nilpotent Lie algebra.
Then $\Lg$ admits a Novikov structure.
\end{prop}

\begin{proof}
It is well known that for $x,y\in \Lg$ the formula $x\cdot y=\frac{1}{2}[x,y]$
defines a left-symmetric structure on $\Lg$. It satisfies $x\cdot(y\cdot z)=0$
for all $x,y,z\in \Lg$. Hence the product is also Novikov.
\end{proof}

If we want to to find an example of a solvable or nilpotent Lie algebra
without any affine structure, we have to solve nonlinear equations
in the entries of the left or right multiplications. This is quite
difficult in general. For a Novikov structure the situation is better.
The following relation follows from the identities in lemma $\ref{nov}$:
\begin{align}
L([x,y])+\ad([x,y]-[\ad(x),L(y)]-[L(x),\ad(y)]& = 0
\end{align}
In fact, we have $0=[R(x),R(y)]=[L(x)-\ad x, L(y)-\ad y]$. Using the fact that
$L$ and $\ad$ are Lie algebra representations the relation follows. It yields {\it linear} equations 
in the entries of the left multiplications which are very helpful for solving the other
non-linear equations. It enables us to prove the non-existence of
Novikov structures on the following $2$-step solvable
Lie algebra:

\begin{ex}
Let  $\Lg$ be the free $4$-step nilpotent Lie algebra on $2$ generators $x_1$
and $x_2$. Let $(x_1,\ldots,x_8)$ be a basis  of $\Lg$ with the following Lie brackets:
\begin{align*} 
x_3 & = [x_1,x_2] \\
x_4 & = [x_1,[x_1,x_2]] = [x_1, x_3]\\
x_5 & = [x_2,[x_1,x_2]] = [x_2,x_3] \\
x_6 & = [x_1,[x_1,[x_1,x_2]]] =  [x_1, x_4] \\
x_7 & = [x_2,[x_1,[x_1,x_2]]] = [x_2,x_4] \\
    & = [x_1,[x_2,[x_1,x_2]]] =  [x_1, x_5] \\
x_8 & = [x_2,[x_2,[x_1,x_2]]] =  [x_2,x_5]
\end{align*}
\end{ex}

\begin{prop}
The  $2$-step solvable Lie algebra from the above example does not admit any 
Novikov structure.
\end{prop}

\begin{proof}
Indeed, the linear equations arising from the above relation, for a Novikov product with this 
Lie algebra, reduce the system of all equations to very few equations -- which are
contradictory. Note that $\Lg$ admits an affine structure since it is positively graded. 
\end{proof}


\section{Construction of Novikov structures via classical r-matrices}

Classical r-matrices arise in the study of the classical Yang-Baxter equation,
differential Lie algebras and Poisson brackets. 
We recall its definition, see \cite{STS}.

\begin{defi}
Let $\Lg$ be a Lie algebra. A linear operator $T \in \End(V)$ over the
vector space of $\Lg$ will be called a classical r-matrix, if the bracket
\begin{align}
[x,y]_T & = [T(x),y]+[x,T(y)]
\end{align}
is a Lie bracket, i.e., satisfies the Jacobi identity.
\end{defi}

This bracket satisfies the Jacobi identity if and only if
\[
[x,[T(y),T(z)]-T([y,z]_T)]]=0
\]
for all $x,y,z \in \Lg$.

\begin{defi}
Let $\Lg$ be a Lie algebra. A linear operator $T \in \End(V)$ satisfies the 
classical Yang-Baxter equation (CYBE) if
\begin{align}\label{CYBE}
[T(x),T(y)] & = T([T(x),y]+[x,T(y)])
\end{align}
for all $x,y \in \Lg$.
\end{defi}

If  $T \in \End(V)$ satisfies the CYBE, then it is also a classical
r-matrix. Hence we obtain two Lie algebra structures on the vector
space $V$: the Lie algebra $(\Lg,[,])$ and the Lie algebra 
$(\Lg_T,[,]_T)$. 
Solutions to the CYBE can be used to construct Novikov
structures on Lie algebras as follows:

\begin{prop}\label{4.2}
Let $\Lg$ be a Lie algebra and $T\in \End(V)$. Suppose that $T$ satisfies
the CYBE and the following identity:
\begin{align}\label{novbed}
[x,T([y,T(z)])] & = [y,T([x,T(z)])]
\end{align}
for all $x,y,z\in \Lg $. Then the product
\begin{align}\label{yangnov}
x\cdot y & = [T(x),y]
\end{align}
is Novikov. It defines a Novikov structure on the Lie algebra
$\Lg_T$.
\end{prop}

\begin{proof}
We already know that the bracket $[x,y]_T$ is a Lie bracket since $T$ satisfies the CYBE.
In fact, we will show that the product \eqref{yangnov} is left-symmetric, so that
we obtain a Lie bracket by forming the commutator:
\[
x\cdot y-y\cdot x=[T(x),y]+[x,T(y)]=[x,y]_T
\]
Denote by $(x,y,z):=(x\cdot y)\cdot z - x\cdot (y\cdot z)$ the associator of three
elements $x,y,z \in \Lg$. Then we have
\begin{align*}
(x,y,z)-(y,x,z) & = [T([T(x),y]),z]-[T(x),[T(y),z]] - [T([T(y),x]),z]+ [T(y),[T(x),z]] \\
 & =  [T([T(x),y]),z] + [T([x,T(y)]),z] - [[T(x),T(y)],z] \\
 & = [0,z]=0
\end{align*}
Hence the product \eqref{yangnov} is left-symmetric. It is also Novikov, since the right
multiplication is given by $R(x)=-\ad x \kringel T$, so that
\begin{align*}
[R(x),R(y)] & = [-\ad x\kringel T, -\ad y\kringel T] \\
 & =  \ad x \kringel T \kringel \ad y \kringel T -  \ad y \kringel T \kringel \ad x \kringel T \\
 & = 0
\end{align*}
by \eqref{novbed}. 
\end{proof}

\begin{rem}
We note that $T \colon \Lg_T \ra \Lg$ is a Lie algebra homomorphism since
\[
T([x,y]_T)= [T(x),T(y)]
\]
\end{rem}

Proposition $\ref{4.2}$ can be used to construct Novikov structures on
certain Lie algebras. In some cases we can easily construct a linear operator $T \in \End(V)$ 
satisfying \eqref{CYBE}, \eqref{novbed}. This yields many examples of Lie algebras
$\Lg_T$ admitting a Novikov structure:

\begin{lem}
Let $\Lg$ be a Lie algebra with basis $(x_1, \ldots , x_n)$, and fix $\ell, m \in \{ 1,\ldots , n\}$.
Define a linear operator $T \in \End(V)$ by
\[
T(x_i) =
\begin{cases}
x_m, & i=\ell \\
0, & i\neq \ell
\end{cases}
\]
Assume that $T([x_i,x_m])=0$ for all $1\le i \le n$. Then $T$ satisfies
\eqref{CYBE}, \eqref{novbed} and defines a Novikov structure on $\Lg_T$.
The Lie brackets of $\Lg_T$ are, for $i<j$, given by
\[
[x_i,x_j]_T =
\begin{cases}
[x_m,x_j] & i=\ell \\
[x_i,x_m], & i\neq \ell ,\, j=\ell\\
0 &  i\neq \ell ,\, j \neq \ell
\end{cases}
\]
\end{lem}

\begin{proof}
We need to verify that $T$ satisfies the identities
\begin{align*}
T([T(x_i),x_j])+T([x_i,T(x_j)]) & = [T(x_i),T(x_j)]\\
[x_i, T([x_j,T(x_k)])] & = [x_j, T([x_i,T(x_k)])]
\end{align*}
for all $i,j,k$. The first identity reduces to $T([T(x_i),x_j])=0$ for $j\neq \ell$.
The latter is clear for $i\neq \ell$ and follows for $i=\ell$ from
$T([x_m,x_j])=0$. The second identity is checked in the same way.
\end{proof}

\begin{ex}
Let $\Lg=\Lr_2(\C)$ be the $2$-dimensional nonabelian Lie algebra with basis $(x_1,x_2)$
and $[x_1,x_2]=x_2$. Then we may define $T \in \End(\C^2)$ by $T(x_1)=x_1$,  $T(x_2)=0$.
Then $T$ satisfies the assumption of the lemma, and hence satisfies \eqref{CYBE}, \eqref{novbed}. 
We obtain a Novikov structure on  $\Lg_T=\Lr_2(\C)$. In fact, $[x_1,x_2]_T=x_2$, so that $\Lg_T=\Lr_2(\C)$. 
\end{ex}

If we have a given Lie algebra $\Lh$, then we would like
to find a pair $(\Lg,T)$ such that $\Lh \cong \Lg_T$ by Proposition $\ref{4.2}$.
In general, such a pair does not exist. Hence the question is which Lie algebras $\Lh$ can be realized 
as a $\Lg_T$. First of all, the structure of  $\Lg_T$ depends on the structure
of $\Lg$. We will show that if $\Lg$ is solvable, then so
is $\Lg_T$. In fact, the solvability degree of $\Lg_T$ cannot be higher than that of $\Lg$:
The same applies for nilpotency. 

\begin{lem}
If we denote by $\Ad x=\ad T(x)+\ad x \kringel T$ the adjoint operator in 
$\End (\Lg_T)$, i.e., $\Ad x (y)=[x,y]_T$, then we have the following identity:
\begin{align*}
\Ad x_1 \kringel \Ad x_2 \kringel \cdots \kringel \Ad x_n & = \ad T(x_1)\kringel \ad T(x_2) \kringel 
\cdots \kringel \ad T(x_n) \\ 
 & + \left( \sum _{i=1}^n \ad T(x_1) \kringel \cdots \kringel \ad x_i \kringel \cdots \kringel \ad T(x_n)  
\right) \circ T
\end{align*}
\end{lem}

\begin{proof}
The formula follows by induction. For $n=1$ we have $\Ad x_1=\ad T(x_1)+\ad x_1 \kringel T$
by definition. The step from $n$ to $n+1$ follows by using
\[
T\kringel \ad T(x_{n+1})+T\kringel \ad x_{n+1}\kringel T = \ad T(x_{n+1})\kringel T
\]
\end{proof}

\begin{cor}
If $\Lg$ is $r$-step nilpotent, then $\Lg_T$ is $s$-step nilpotent with $s\le r$.  
\end{cor}

\begin{proof}
By assumption, $\ad x_1 \kringel \cdots \kringel \ad x_r=0$ for all $x_1,\ldots ,x_r \in \Lg$.
Then the formula in the above lemma implies $\Ad x_1 \kringel \cdots \kringel \Ad x_r=0$.
\end{proof}

\begin{cor}
If $\Lg$ is $r$-step solvable, then $\Lg_T$ is $s$-step solvable with $s\le r$.
\end{cor}

\begin{proof}
This follows again from the formula. We leave this to the reader. Instead we show 
directly that $\Lg_T^{(n)}\subseteq \Lg^{(n)}$ for all $n\ge 1$. 
The case $n=1$ is obvious. Assume that $\Lg_T^{(n-1)}\subseteq \Lg^{(n-1)}$. 
The space $\Lg_T^{(n)}$ is spanned by elements of the form $[x,y]_T$ with $x,y\in \Lg_T^{(n-1)}$.
We have $T(x), T(y) \in  \Lg^{(n-1)}$, so that
\[
[x,y]_T = [T(x),y]+[x,T(y)] \in [ \Lg^{(n-1)}, \Lg^{(n-1)}]= \Lg^{(n)}.
\]
\end{proof}

As an application we construct Novikov structures via Proposition $\ref{4.2}$
on $3$-dimensional complex Lie algebras, which are listed below:

\vspace*{0.5cm}
\begin{center}
\begin{tabular}{c|c}
 $\Lg$ & Lie brackets  \\
\hline
$\C^3$ & $-$ \\
$\Ln_3(\C)$ & $[e_1,e_2]=e_3$ \\
$\Lr_3(\C)$ & $[e_1,e_2]=e_2,\, [e_1,e_3]=e_2+ e_3$ \\
$\Lr_{3,\la}(\C)$ & $[e_1,e_2]=e_2, \,[e_1,e_3]=\la e_3$ \\
$\Ls \Ll_2 (\C)$ & $[e_1,e_2]=e_3, [e_1,e_3]=-2 e_1,[e_2,e_3]=2 e_2$ 
\end{tabular}
\end{center}
\vspace*{0.5cm}

\begin{prop}
Let $\Lg=\Ls \Ll_2 (\C)$. Assume that $T$ satisfies  \eqref{CYBE}, \eqref{novbed}. Then
$\Lg_T$ is isomorphic to one of the following Lie algebras: $\C^3,\, \Ln_3(\C)$ or
$\Lr_{3,-1}(\C)$. All three Lie algebras arise as a $\Lg_T$ for a suitable $T$.
\end{prop}

\begin{proof}
Let $E_{ij}$ be the matrix in $M_3(\C)$ with a $1$ at the place $(i,j)$ and zeros
otherwise. For $T=0$ we have $\Lg_T=\C^3$. For $T=E_{12}$ we have  $\Lg_T=\Ln_3(\C)$.
Finally, for $T=E_{33}$ we have $\Lg_T \cong \Lr_{3,-1}(\C)$.
Conversely a direct calculation shows that the Lie algebra $\Lg_T$ for any such $T$
must be isomorphic to one of these Lie algebras. In fact, suppose that $T=(a_{ij})$.
If $a_{12}\neq 0$ we may assume that $a_{12}=1$. Let $\al= a_{11}$ and $\be =a_{32}$. The
conditions \eqref{CYBE}, \eqref{novbed} for $T$ imply that
\[
T= \begin{pmatrix} \al & 1 & 2\be\\  \al^2 & \al & 2 \al\be 
\\ \al\be & \be & 2\be^2 \end{pmatrix},\quad
\]
If $\al+\be^2=0$, then $\Lg_T$ is $2$-step nilpotent, hence isomorphic to $\Ln_3(\C)$.
Otherwise $\Lg_T$ is solvable, not nilpotent, unimodular and 
isomorphic to $\Lr_{3,-1}(\C)$. The case  $a_{12}= 0$ is similar.
\end{proof}

In the same way we obtain:

\begin{prop}
For $\Lg=\Lr_{3,-1}(\C)$, $\Lg_T$ is isomorphic to $\C^3,\, \Ln_3(\C)$ or $\Lr_{3,\la}(\C)$. 
For $\Lg=\Lr_{3}(\C)$, $\Lg_T$ is isomorphic to $\C^3,\, \Ln_3(\C)$ or $\Lr_{3}(\C)$. 
For $\Lg=\Ln_3(C)$,  $\Lg_T$ is isomorphic to $\C^3$ or $\Ln_{3}(\C)$.
All cases can be realized as a $\Lg_T$ with a suitable $T$. 
\end{prop}

\begin{rem}
We have equipped that way all solvable complex $3$-dimensional Lie algebras with a Novikov structure.
\end{rem}

\section{Construction of Novikov structures via extensions}

In the following we will consider Lie algebras $\Lg$ which are an extension
of a Lie algebra $\Lb$ by an abelian Lie algebra $\La$. Hence we have a
short exact sequence of Lie algebras
\begin{equation*}
0 \rightarrow \La \xrightarrow{\iota} \Lg \xrightarrow{\pi}
\Lb \rightarrow 0
\end{equation*}
Since $\La$ is abelian, there exists a natural $\Lb$-module structure
on $\La$. We denote the action of $\Lb$ on $\La$ by $(x,a)\mapsto \phi (x)a$,
where $\phi \colon \Lb \ra \End (\La)$ is the corresponding Lie algebra
representation. We have
\begin{equation}
\phi([x,y])=\phi(x)\phi(y)-\phi(y)\phi(x)\label{5}
\end{equation}
for all $x,y \in \Lb$.
Let $\Om \in Z^2(\Lb,\La)$ be a $2$-cocycle. This means that
$\Om : \Lb \times \Lb \ra \La$ is a skew-symmetric bilinear map
satisfying
\begin{equation}\label{6}
\phi(x)\Om(y,z)-\phi(y)\Om(x,z)+\phi(z)\Om (x,y) =\Om ([x,y],z)-\Om ([x,z],y)+\Om([y,z],x)
\end{equation}
We obtain a Lie bracket on $\Lg=\La \times \Lb$ by
 \begin{equation}\label{lie-algebra}
[(a,x),(b,y)]:=(\phi (x)b-\phi(y)a+\Om(x,y),[x,y])
\end{equation}
for $a,b\in \La$ and $x,y\in \Lb$.
As a special case we obtain the $2$-step solvable Lie algebras as extensions
of two abelian Lie algebras $\La$ and $\Lb$:

\begin{rem}
Any two-step solvable Lie algebra $\Lg$ is described by the following
data: there is an exact sequence of Lie algebras
\begin{equation*}
0 \rightarrow \La \xrightarrow{\iota} \Lg \xrightarrow{\pi}
\Lb \rightarrow 0
\end{equation*}
such that: $\La$ and $\Lb$ are abelian Lie algebras,
$\phi : \Lb \ra \End (\La)$ is a Lie algebra representation, 
$\Om \in Z^2(\Lb,\La)$ is a $2$-cocycle, and the Lie bracket of
$\Lg=\La \times \Lb$ is given by
 \begin{equation*}
[(a,x),(b,y)]:=(\phi (x)b-\phi(y)a+\Om(x,y),0)
\end{equation*}
\end{rem}

Suppose we are given such an extension. Then $\La$ and $\Lb$ are vector
spaces with trivial Lie brackets. In this case the 
Lie algebra $\Lg$ is clearly two-step solvable.
The conditions on $\phi$ and $\Om$ reduce as follows: since $\La$ and $\Lb$
are abelian, $\phi$ is just a linear map satisfying
\begin{equation*}
\phi(x)\phi(y)=\phi(y)\phi(x)
\end{equation*}
for all $x,y \in \Lb$.
On the other hand, $\Om : \Lb \times \Lb \ra \La$
is a skew-symmetric bilinear map satisfying
\begin{equation*}
\phi(x)\Om(y,z)-\phi(y)\Om(x,z)+\phi(z)\Om (x,y) =0
\end{equation*}

Conversely, let $\Lg$ be a two-step solvable Lie algebra.
Then we have an extension of Lie algebras
\begin{equation*}
0 \rightarrow \Lg^{(2)} \xrightarrow{\iota} \Lg \xrightarrow{\pi}
\Lg/\Lg^{(2)} \rightarrow 0
\end{equation*}
where $\La=\Lg^{(2)}$ and $\Lb=\Lg/\Lg^{(2)}$ are
abelian Lie algebras. In that case, $\Lg$ acts on $\La$
by the adjoint representation and $\phi$ corresponds to the induced action
of $\Lg/\La$ on $\La$. The skew-symmetric bilinear map $\Om$ is given by
$\Om(x,y)=\iota^{-1}([\tau(x),\tau(y)]-\tau([x,y]))=
\iota^{-1}([\tau(x),\tau(y)])$,
for $x,y\in \Lb$, where $\tau$ is a section, i.e., a linear map $\tau : \Lb
\ra \Lg$ satisfying $\pi\circ\tau=\id$ on $\Lb$. \\[0.3cm]
Now we want to construct affine structures on Lie algebras $\Lg$ which are
an extension of a Lie algebra $\Lb$ by an abelian Lie algebra $\La$.
Assume that $\Lg=(\La,\Lb,\phi,\Om)$ is an extension with the above data.
Suppose that we have already an LSA-product  $(a,b)\mapsto a\cdot b$ on
$\La$ and an LSA-product $(x,y)\mapsto x\cdot y$ on $\Lb$.
Since  $\La$ is abelian, the LSA-product on $\La$ is commutative and
associative. Hence it defines also a Novikov structure on $\La$.
We want to lift these  LSA-products to $\Lg$. Consider
\begin{align*}
\om & \colon \Lb \times \Lb \ra \La\\
\phi_1,\, \phi_2 & \colon \Lb \ra\End (\La)
\end{align*}
where $\om$ is a bilinear map and 
$\phi_1,\, \phi_2$ are Lie algebra representations.  
We will define a bilinear product $\Lg \times \Lg \ra \Lg$ by
\begin{equation}\label{produkt}
(a,x)\kringel (b,y):=(a\cdot b+\phi_1(y)a+\phi_2 (x)b+\om(x,y),x\cdot y)
\end{equation}  

\begin{prop}\label{phi12}
The above product defines a left-symmetric structure on $\Lg$ 
if and only if the following conditions hold:
\begin{align}
\om(x,y)-\om(y,x) & = \Om(x,y) \label{8}\\
\phi_2(x)-\phi_1(x) & = \phi(x)\label{9}\\
\phi_2(x)\om(y,z)- \phi_2(y)\om(x,z) - \phi_1(z)\Om(x,y) & =
\om(y,x\cdot z)-\om (x,y\cdot z) +\om ([x,y],z)\label{10}\\
a\cdot \om(y,z)+\phi_1 (y\cdot z)a & = \phi_2(y)\phi_1(z)a-\phi_1(z)\phi(y)a \label{11} \\
a\cdot (\phi_1(z)b) & = b\cdot (\phi_1(z)a) \label{12}\\
\phi_2(y)(a\cdot c)- a\cdot (\phi_2(y)c) & = (\phi(y)a)\cdot c \label{13} \\
\Om(x,y)\cdot c & =0\label{14}
\end{align}
for all $a,b,c \in \La$ and  $x,y,z \in \Lb$.
\end{prop}

\begin{proof}
Let $u=(a,x), v=(b,y), w=(c,z)$ denote three arbitrary elements of $\Lg$. 
Let us first consider the equation \eqref{lsa2} for the product, i.e.,
\begin{align*}
[u,v] & =u\kringel v - v\kringel u
\end{align*}
Using \eqref{lie-algebra}, \eqref{produkt} and the commutativity of
the LSA-product in $\La$ we obtain
\begin{align*}
[u,v] & =(\phi(x)b-\phi(y)a+\Om(x,y),[x,y])\\
u\kringel v -v\kringel u& =((\phi_2(x)-\phi_1(x))b-(\phi_2(y)-\phi_1(y))a
+\om(x,y)-\om(y,x),[x,y])
\end{align*}
Suppose that the two expressions are equal for all $a,b\in\La$ and
$x,y\in \Lb$. For $a=b=0$ we obtain $\om(x,y)-\om(y,x)=\Om(x,y)$. Taking this into
account, $a=0$ implies $\phi_2(x)-\phi_1(x)= \phi(x)$. Conversely, these
two conditions imply \eqref{lsa2}.\\
The computation required for the left-symmetric condition \eqref{lsa1}
is a bit longer:

\begin{align*}
\begin{split}
u\kringel (v\kringel w)-v\kringel (u\kringel w) & =\bigl( a\cdot(b\cdot c)-
b\cdot(a\cdot c)+a\cdot(\phi_1(z)b)-b\cdot(\phi_1(z)a) \\
& +a\cdot(\phi_2(y)c)- b\cdot(\phi_2(x)c) + a\cdot\om(y,z)-b\cdot \om(x,z) \\
& +\phi_1(y\cdot z)a- \phi_1(x\cdot z)b+\phi_2(x)(b\cdot c)-\phi_2(y)(a\cdot c)\\
& + \phi_2(x)\phi_1(z)b-\phi_2(y)\phi_1(z)a+[\phi_2(x),\phi_2(y)]c
 + \phi_2(x)\om(y,z)\\
& -\phi_2(y)\om(x,z)+\om(x,y\cdot z)-\om(y,x\cdot z),\; x\cdot(y\cdot z)-
y\cdot (x\cdot z)\bigr) \\[0.2cm]
[u,v]\kringel w & = ((\phi(x)b)\cdot c-(\phi(y)a)\cdot c+\Om (x,y)\cdot c
+\phi_1(z)\phi(x)b \\
& - \phi_1(z)\phi(y)a + \phi_1(z)\Om(x,y)+\phi_2([x,y])c+\om([x,y],z),\; [x,y]\cdot z )\\
\end{split}
\end{align*}

The product is left-symmetric, i.e., 
condition \eqref{lsa1} is satisfied, if and only if these two expressions
are always equal.
For $a=b=c=0$ this yields condition \eqref{10}. Using this,
and setting $a=b=0$, one obtains \eqref{14}. 
Again using this and letting $b=c=0$, one obtains \eqref{11}.
Taking $z=x=b=0$ yields \eqref{13}. Finally the choice of $c=x=y=0$
gives \eqref{12}. 
Conversely, the above conditions ensure that condition \eqref{lsa1} is 
satisfied.   
\end{proof}  

\begin{prop}\label{novikov}
The product \eqref{produkt} defines a Novikov structure on $\Lg$ if and
only if the conditions for a left-symmetric structure are satisfied, and
in addition the following conditions hold:
\begin{align}
\phi_1(z)\om(x,y)-\phi_1(y)\om(x,z) & = \om(x\cdot z,y)-\om(x\cdot y,z) \label{15}\\
\om(x,y)\cdot c+\phi_2(x\cdot y)c & = \phi_1(y)\phi_2(x)c \label{16}\\
[\phi_1(x),\phi_1(y)] & = 0 \label{17}\\
(\phi_2(x)b)\cdot c & = (\phi_2(x)c)\cdot b \label{18}\\
\phi_1(z)(a\cdot b) & = (\phi_1(z)a)\cdot b \label{19}\\
(x\cdot y)\cdot z & = (x\cdot z)\cdot y \label{20} 
\end{align}
\end{prop}

\begin{proof}
Since the product on $\La$ is associative and commutative we have automatically
\begin{align}
(a\cdot b)\cdot c & = (a\cdot c)\cdot b \label{21}
\end{align}
Let $u=(a,x), v=(b,y), w=(c,z)$ as before. We have

\begin{align*}
\begin{split}
(u\kringel v)\kringel w & = ((a\cdot b)\cdot c + (\phi_1(y)a)\cdot c
+(\phi_2(x)b)\cdot c + \om(x,y)\cdot c \\
& + \phi_1(z)(a\cdot b) + \phi_1(z)\phi_1(y)a + \phi_1(z)\phi_2(x)b \\
& + \phi_1(z)\om (x,y)+\phi_2(x\cdot y)c+\om (x\cdot y,z),\; (x\cdot y)\cdot z)\\[0.3cm]
(u\kringel w)\kringel v & = ((a\cdot c)\cdot b + (\phi_1(z)a)\cdot b
+(\phi_2(x)c)\cdot b + \om(x,z)\cdot b \\
& + \phi_1(y)(a\cdot c) + \phi_1(y)\phi_1(z)a + \phi_1(y)\phi_2(x)c \\
& + \phi_1(y)\om (x,z)+\phi_2(x\cdot z)b+\om (x\cdot z,y),\; (x\cdot z)\cdot y) 
\end{split}
\end{align*}

Suppose that the product is Novikov, i.e., 
suppose that $(u\kringel v)\kringel w= 
(u\kringel w)\kringel v$. Then for $a=b=c=0$ we obtain the conditions \eqref{15} and
\eqref{20}. For $x=y=z=0$ we obtain \eqref{21}, and for $a=b=0$ it follows
\eqref{16}. For $b=c=0$ it follows \eqref{17} and for $y=z=a=0$ we obtain
\eqref{18}. Finally setting $x=y=c=0$ yields \eqref{19}. 
Conversely, the above conditions ensure that the product is Novikov. 
\end{proof}

If $\Lb$ is also abelian, then the Lie algebra $\Lg$ is two-step solvable.
We say that the LSA-products on $\La$ and $\Lb$ are trivial if
$a\cdot b=x\cdot y=0$ for all $a,b\in \La$ and $x,y\in \Lb$.

\begin{cor}\label{trivial}
Suppose that the LSA-products on $\La$ and $\Lb$ are trivial.
Hence $\Lb$ is also abelian.
Then \eqref{produkt} defines a left-symmetric structure
on $\Lg$ if and only if the following conditions hold:

\begin{align*}
\om(x,y)-\om(y,x) & = \Om(x,y)\\
\phi_2(x)-\phi_1(x) & = \phi(x)\\
\phi_2(x)\om(y,z)- \phi_2(y)\om(x,z)& = \phi_1(z)\Om(x,y)\\
[\phi_1(x),\phi_2(y)] & = \phi_1(x)\phi_1(y)
\end{align*}

It defines a Novikov structure on $\Lg$ if in addition
\begin{align*}
\phi_1(z)\om(x,y) & = \phi_1(y)\om(x,z)\\
\phi_1(x)\phi_2(y) & = 0\\
[\phi_1(x),\phi_1(y)] & = 0
\end{align*}
\end{cor}

In particular, if $\phi_1=0$, the product defines a Novikov
structure on $\Lg$ if and only if it defines a left-symmetric structure
on $\Lg$.

\begin{cor}
Assume that $\Lg= \La \rtimes_{\phi} \Lb$ is a semidirect product of an abelian
Lie algebra $\La$ and a Lie algebra $\Lb$ by a representation 
$\phi\colon \Lb \ra \End(\La)=\Der(\La)$. This yields a split exact sequence
\begin{equation*}
0 \rightarrow \La \xrightarrow{\iota} \Lg \xrightarrow{\pi}
\Lb \rightarrow 0
\end{equation*}
If $\Lb$ admits an LSA-product then also $\Lg$ admits an LSA-product.
If $\Lb$ admits a Novikov product $(x,y)\mapsto x\cdot y$ such that
\[
\phi(x\cdot y)=0
\] 
for all $x,y\in \Lb$ then also $\Lg$ admits a Novikov product.
\end{cor}

\begin{proof}
Because the short exact sequence is split, the $2$-cocycle $\Om$ in the Lie bracket
of $\Lg$ is trivial, i.e., $\Om(x,y)=0$. Let $a\cdot b=0$ be the trivial product
on $\La$ and take $\phi_1=0$, $\om(x,y)=0$. Assume that $(x,y)\mapsto x\cdot y$ is
an LSA-product. Then all conditions of Proposition $\ref{phi12}$ are satisfied.
Hence \eqref{produkt} defines an LSA-product on $\Lg$, given by
\[
(a,x)\kringel (b,y)=(\phi(x)b, x\cdot y)
\]
Assume that the product on $\Lb$ is Novikov. Then the product on $\Lg$ will be Novikov
if and only if the conditions of Proposition $\ref{novikov}$ are satisfied.
In this case, only condition \eqref{16} remains, which says $\phi(x\cdot y)=0$. 
\end{proof}

\begin{ex}\label{ex-3.5}
Let $\Lg$ be the $5$-dimensional Lie algebra with basis $(A,B,C,X,Y)$
and Lie brackets $[X,Y]=A,\; [X,A]=B,\; [Y,A]=C$. This is a free 
$3$-step nilpotent Lie algebra. Hence it is $2$-step solvable.
The product \eqref{produkt} defines a Novikov structure on $\Lg$ with 

\begin{align*}
\phi_1(X) & = \begin{pmatrix} 0 & 0 & 0\\ -1/2 & 0 & 0\\ 0 & 0 & 0 \end{pmatrix},\quad
\phi_1(Y)=\begin{pmatrix} 0 & 0 & 0\\ 0 & 0 & 0\\ 0 & 0 & 0 \end{pmatrix} \\
\phi_2(X) & = \begin{pmatrix} 0 & 0 & 0\\ 1/2 & 0 & 0\\ 0 & 0 & 0 \end{pmatrix},\quad
\phi_2(Y)= \begin{pmatrix} 0 & 0 & 0\\ 0 & 0 & 0\\ 1 & 0 & 0 \end{pmatrix} \\
\end{align*}
and $\om (Y,X)=-A,\; \om(X,X)= \om(X,Y)= \om(Y,Y)=0$, and trivial products on
$\La$ and $\Lb$.
\end{ex}

The Lie algebra $\Lg$ is given by
$\Lg=(\La,\Lb,\phi,\Om)$ with $\La=\s \{A,B,C\},\; \Lb=\s \{X,Y\}$ abelian,
$\Om(X,Y)=A$ and $\phi(X)A=B,\;\phi(Y)A=C $, i.e.,
$$
\phi(X) = \begin{pmatrix} 0 & 0 & 0\\ 1 & 0 & 0\\ 0 & 0 & 0 \end{pmatrix},\quad
\phi(Y) = \begin{pmatrix} 0 & 0 & 0\\ 0 & 0 & 0\\ 1 & 0 & 0 \end{pmatrix}
$$

It is easy to see that the conditions of corollary $\ref{trivial}$ are satisfied.
Note that any product $\phi_i(x)\phi_j(y)=0$ for $1\le i,j\le 2$ and
$x,y\in \Lb$. We can write down the resulting Novikov structure on
$\Lg$ explicitely. It is given by \eqref{produkt}; we only write the non-zero
products:

\begin{align*}
A\kringel X & = -B/2\\
X\kringel A & = B/2\\
Y\kringel A & = C\\
Y\kringel X & = -A
\end{align*}

\begin{lem}
Let $\Lg=(\La,\Lb,\phi,\Om)$ be a two-step solvable Lie algebra
equipped with a left-symmetric structure comming from \eqref{produkt}.
Then $\La$ is a two-sided ideal in the left-symmetric algebra $\Lg$.
\end{lem}

\begin{proof}
$\La$ is isomorphic to $\{(a,0)\in \La\times\Lb=\Lg \mid a\in \La\}$. We have
\begin{align*}
(a,0)\kringel (b,y) & = (a\cdot b +\phi_1(y)a,0) \in \La \\
(b,y)\kringel (a,0) & = (b\cdot a+\phi_2(y)a,0) \in \La
\end{align*}
\end{proof}

\begin{rem}
The lemma shows that there are two-step solvable, non-nilpotent Lie algebras admitting
left-symmetric structures which do not arise by \eqref{produkt}.
To see this take, for example, any simple LSA with a two-step
solvable Lie algebra. Note however, that the Lie algebra of a simple LSA cannot 
be nilpotent. There are many examples, see \cite{BU2}, i.e.,
consider for any $n\ge 2$ the algebra
$$
I_n=\langle e_1,\ldots ,e_n \mid e_1\kringel e_1=2e_1,\, e_1\kringel
e_j=e_j,\, e_j\kringel e_j=e_1, \; j\ge 2 \rangle
$$
This is a simple LSA. Hence it does not admit a non-trivial two-sided
ideal $\La$. Its Lie algebra $\Lg$ is given by $[e_1,e_j]=e_j$ for
$j=2,3,\ldots ,n$. It is two-step solvable and the LSA-structure on
it cannot arise by \eqref{produkt}. Note that, of course, it does
nevertheless admit some Novikov structure. Just consider the structure
given by $ e_1\kringel e_j=e_j$ for all $j\ge 2$, and the
other products equal to zero.
\end{rem}

\begin{prop}\label{iso}
Let $\Lg=(\La,\Lb,\phi,\Om)$ be a two-step solvable Lie algebra.
If there exists an $e\in \Lb$ such that $\phi(e)\in\End(\La)$ is an
isomorphism, then $\Lg$ admits a Novikov structure. In fact, in that
case \eqref{produkt} defines a Novikov product, where $\phi_1=0,
\, \phi_2=\phi$, the product on $\La$ and $\Lb$ is trivial, and
\begin{align*}
\om(x,y) & = \phi(e)^{-1}\phi(x)\Om(e,y)
\end{align*}
\end{prop}

\begin{proof}
We have to show that the conditions of corollary $\ref{trivial}$ are satisfied.
Applying $\phi(e)^{-1}$ to \eqref{6} with $z=e$ it follows
$\Om(x,y)-\phi(e)^{-1}\phi(x)\Om(e,y)+\phi(e)^{-1}\phi(y)\Om(e,x)=0$. This
just means that $\Om(x,y)=\om(x,y)-\om(y,x)$.
Furthermore we have
\begin{align*}
\phi(x)\om(y,z)-\phi(y)\om(x,z) & = \phi(x)\phi(e)^{-1}\phi(y)\Om(e,z)-
\phi(y)\phi(e)^{-1}\phi(x)\Om(e,z) \\
 & = \phi(e)^{-1}(\phi(x)\phi(y)-\phi(y)\phi(x))\Om(e,z)\\
 & = 0
\end{align*}
Hence the product defines a left-symmetric structure.
Since $\phi_1=0$ the structure is also Novikov.
\end{proof}

We can apply our results to give a new proof of Scheuneman's result \cite{SCH}:
\begin{prop}\label{scheuneman}
Any three-step nilpotent Lie algebra admits a LSA structure.
\end{prop}

\begin{proof}
Let $\Lg$ be a three-step nilpotent Lie algebra. Since $\Lg$
is two-step solvable, we can use Proposition $\ref{phi12}$.
We will show the existence of a left-symmetric structure
on $\Lg$ as follows.
Let $(e_1,\dots, e_m)$ be a basis of $\Lb$,
$(f_1,\dots ,f_n)$ be a basis of $\La$ and introduce the
following notation.
\begin{align*}
X_i & = \phi_1(e_i) \\
Y_i & = \phi_2(e_i) \\
A_i & = \phi(e_i) \\
x_{ij} & = \om (e_i,e_j)\\
v_{ij} & = \Om (e_i,e_j)
\end{align*} 
for $1\le i,j\le m$. The Lie algebra $\Lg$ is determined by the data
$\La, \Lb, A_i, v_{ij}$ satisfying
\begin{align*}
[A_i,A_j] & = 0\\
v_{ji} + v_{ij} & = 0\\
A_iv_{jk}-A_jv_{ik}+A_kv_{ij} & =0.\\
\end{align*} 
We have $\La=[\Lg,\Lg]$ and $\Lb=\Lg/[\Lg,\Lg]$. The operators
$A_i,A_j\in \End(\La)$ satisfy
$A_iA_j([e_k,e_l])=[e_i,[e_j,[e_k,e_l]]]=0$ since $\Lg$ is
$3$-step nilpotent. Thus we have 
\begin{align*}
A_iA_j & =0.
\end{align*}
Let $Y_i:=X_i+A_i$.
We have to find $x_{ij}\in \La$ and $X_i$ such that
\begin{align*}
Y_ix_{jk}-Y_jx_{ik}-X_k v_{ij} & =0\\
x_{ij}-x_{ji} & = v_{ij}\\
[X_i,A_j] & =X_jX_i
\end{align*}
for all $i,j$. 
A possible solution is given by 
\begin{align*}
x_{ij} & = \frac{1}{2}v_{ij}\\
X_i & = -\frac{1}{3}A_i \\
Y_i & = \frac{2}{3}A_i 
\end{align*}
Then the first equation is just
$\frac{1}{3}(A_iv_{jk}-A_jv_{ik}+A_kv_{ij})=0$,
which is true by assumption. The third equation is trivially
true since $A_jA_i=0$ for all $i,j$. 
\end{proof}

\begin{rem}
It is natural to ask whether the above result holds also for Novikov
structures. We will show that all $3$-step nilpotent Lie algebras with
$2$ or $3$ generators admit a Novikov structure. However, a Novikov structure
may not exist in general on a $3$-step nilpotent Lie algebra. Unfortunately
it seems to be hard to find a counter-example because of dimension reasons.
We will later generalize Scheuneman's result, see proposition \eqref{prop5.7}.
\end{rem}

Suppose that $\Lg=(\La,\Lb,\phi,\Om)$ is a two-step solvable Lie algebra
with the notations as above. 
The assumptions on $A_i$ and $v_{ij}$ for $1\le i,j \le m$
are as follows, see \eqref{5},\eqref{6}:

\begin{align}
v_{ij}+v_{ji} & = 0 \label{22}\\
A_iA_j -A_jA_i & = 0 \label{23}\\
A_iv_{jk}-A_jv_{ik}+A_kv_{ij} & = 0 \label{24}
\end{align}

If we assume that the products on $\La$ and $\Lb$ are trivial, then 
the conditions of corollary $\ref{trivial}$ mean that we 
have to find, for a Novikov structure, $X_i$ and $x_{ij}$ such that,
for all $i,j,k$

\begin{align}
x_{ij}-x_{ji} & = v_{ij} \label{25}\\
X_i +A_i & = Y_i \label{26} \\
Y_ix_{jk}-Y_jx_{ik} & =X_k v_{ij} \label{27} \\
X_iA_j-A_jX_i & = X_jX_i \label{28} \\
X_kx_{ij} & = X_jx_{ik}\label{29} \\
X_iY_j & = 0 \label{30} \\
X_iX_j & = X_jX_i \label{31}
\end{align}

Note that \eqref{23},\eqref{26},\eqref{28} imply 
$$
Y_iY_j=Y_jY_i
$$

Indeed,
\begin{align*}
[Y_i,Y_j] & =[X_i+A_i,X_j+A_j]=[X_i,X_j]+[X_i,A_j]+[A_i,X_j]+[A_i,A_j] \\
 & = [X_i,X_j]+X_jX_i-X_iX_j =0.
\end{align*}

\begin{prop}
Let $\Lg=(\La,\Lb,\phi,\Om)$ be a $2$-generated $3$-step nilpotent
Lie algebra, i.e., $\dim \Lb=2$. Then $\Lg$ admits a Novikov structure.
\end{prop}

\begin{proof}
A $2$-generated $3$-step nilpotent Lie Algebra is a quotient of the $5$-dimensional
free nilpotent Lie algebra of example \eqref{ex-3.5}. It would not be difficult
to verify directly that such an algebra admits a Novikov structure. 
However in this case we can simply use corollary $\ref{trivial}$ to construct a Novikov structure.
Let $\Lb=\langle e_1,e_2 \rangle$. Note that the assumption \eqref{24} is automatically satisfied since
$\dim \Lb=2$: there are always two indices equal, so that any skew-symmetric map
$\Om$ is a $2$-cocycle.  
We have to find $X_1,X_2\in \End(\La)$ and $x_{ij}\in \La$
such that the conditions \eqref{25},\ldots ,\eqref{31} hold.
It is easy to see that a solution is given by
\begin{align*}
X_1 & = -A_1/2 \\
X_2 & = 0 \\
x_{11} & = 0, \; x_{12}  = 0, \; x_{21} = -v_{12}, \; x_{22} = 0 
\end{align*}
\end{proof}

\begin{ex}
Let $\Lf$ be the free $3$-step nilpotent Lie algebra with $3$ generators. Then
$\Lf$ admits a Novikov structure.
\end{ex}
Let $(x_1,\ldots ,x_{14})$ be a basis of $\Lf$ with generators $(x_1,x_2,x_3)$ and Lie brackets

\begin{align*}
x_4 & = [x_1,x_2] \\
x_5 & = [x_1,x_3] \\
x_6 & = [x_2,x_3] \\
x_7 & = [x_1,[x_1,x_2]] = [x_1,x_4]\\
x_8 & = [x_2,[x_1,x_2]] = [x_2,x_4]\\
x_9 & = [x_3,[x_1,x_2]] = [x_3,x_4]\\
x_{10} & = [x_1,[x_1,x_3]] = [x_1,x_5]\\
x_{11} & = [x_2,[x_1,x_3]] = [x_2,x_5]\\
x_{12} & = [x_3,[x_1,x_3]] = [x_3,x_5]\\
x_{11}-x_9 & = [x_1,[x_2,x_3]] = [x_1,x_6]\\
x_{13} & = [x_2,[x_2,x_3]] = [x_2,x_6]\\
x_{14} & = [x_3,[x_2,x_3]] = [x_3,x_6]\\
\end{align*}

The Jacobi identity is satisfied. A Novikov structure is given by 

\begin{align*}
x_1\kringel x_3 & = x_5,\quad x_1\kringel x_4=\frac{1}{2}x_7,\quad x_1\kringel x_5 = x_{10},
\quad x_1\kringel x_6 = x_{11}-\frac{1}{2}x_9 \\
x_2\kringel x_1 & = -x_4,\quad  x_2\kringel x_3 = x_6,\quad  x_2\kringel x_4 = x_8,\quad
x_2\kringel x_5 = x_{11},\quad   x_2\kringel x_6 = x_{13}\\
x_3\kringel x_4 & = \frac{1}{2}x_9,\quad x_3\kringel x_5  = \frac{1}{2}x_{12},\quad
x_3\kringel x_6  = \frac{1}{2}x_{14},\quad x_4\kringel x_1  = -\frac{1}{2}x_{7},\\
x_4\kringel x_3  & = -\frac{1}{2}x_{9}, \quad x_5\kringel x_3  = -\frac{1}{2}x_{12},\quad
x_6\kringel x_1  = \frac{1}{2}x_{9},\quad x_6\kringel x_3  = -\frac{1}{2}x_{14}
\end{align*}
\vspace{0.1cm}
\begin{prop} Any  3-generated 3-step nilpotent Lie algebra admits a Novikov structure.
\end{prop}
\begin{proof} Let $\Lg$ be a 3-generated 3-step nilpotent Lie algebra. Then $\Lg$ is isomorphic to a
quotient $\Lf/I$, where $\Lf$ denotes the free 3-step nilpotent Lie algebra on 3 generators and 
$I$ is a Lie algebra ideal in $\Lf$. In the example above we showed that $\Lf$ admits a Novikov 
structure. To prove this proposition it is enough to show that we can choose $I$ in such a way that 
it is also a 2-sided ideal for the Novikov-product. For in this situation, the product on $\Lf$ is 
inherited by $\Lg$.
We can distinguish three cases:
\begin{itemize}
\item[(1)] $\dim \Lg^2/\Lg^3=3$: In this case $I\subseteq Z(\Lf)$ 
and it is obvious that $I$ is an ideal for the Novikov product. 
\item[(2)] $\dim \Lg^2/\Lg^3=2$: This is only possible if $I=\langle c+z \rangle \oplus Z$, where 
$c\in \Lf^2$, $z\in \Lf^3=Z(\Lf)$ and $Z$ is a subspace of $Z(\Lf)$. As $\Lf$ is generated by 3 elements,
any element of $\Lf^2$ can be written as a genuine Lie bracket $c=[u,v]$ for some $u,v\in \Lf$ (and not 
only as a linear combination of Lie brackets). Therefore, we can without loss of generality assume that, 
with the notations from the example  above, $c=[x_1,x_2]=x_4$. Since $I$ is a Lie algebra ideal, 
we obtain that $[x_1, c+z]=[x_1,x_4]=x_7\in I,\;[x_2,c+z]=x_8\in I \mbox{ and }
[x_3,c+z]=x_9 \in I.$
It follows that $I=\langle x_4 +z, x_7,x_8,x_9\rangle \oplus Z'$, with $Z'\subseteq \Lf^3$. It is now obvious again
that $I$ is actually a Novikov ideal.
\item[(3)] $\dim \Lg^2/\Lg^3=1$: Analogously as in the previous case, we can assume that 
$I=\langle x_4+z_1, x_5+z_2 \rangle  \oplus Z$, where $z_1,z_2\in \Lf^3$ and $Z\subseteq \Lf^3$ and 
continue exactly as before.
\end{itemize}
\end{proof}

Let $J(n)$ denote the full Jordan block of size $n$ to the eigenvalue zero, i.e.,
$$
J(n)={\Small
\begin{pmatrix}
 0 & 1 & \hdots & 0 & 0\\
 0 & 0 & \hdots & 0 & 0\\
\vdots & \vdots & \ddots & \vdots & \vdots\\
0 & 0  & \hdots & 0 & 1\\
0 & 0  & \hdots & 0 & 0
\end{pmatrix}}
$$

\begin{prop}
Let $\Lg=(\La,\Lb,\phi,\Om)$ be a two-step solvable Lie algebra.
If there exists an $x\in \Lb$ such that the Jordan form of $\phi(x)\in\End(\La)$ 
is $J(n)$, where $\dim \La=n$, then $\Lg$ admits a Novikov structure. 
\end{prop}

\begin{proof}
We may assume that $(e_1,\ldots ,e_m)$ is a basis for $\Lb$ such that
$A_1=\phi (e_1)=J(n)$. Since all $A_i$ commute with $A_1$ by \eqref{23}
we have
\begin{align}
A_i & = \ga_{i,0}E +\ga_{i,1}A_1+ \ga_{i,2}A_1^2+ \cdots + \ga_{i,n-1}A_1^{n-1}
\end{align}
for $\ga_{i,j} \in k$. If $\ga_{i,0} \ne 0$ for some $i$ then it follows
$\det A_i\ne 0$. Then $\Lg$ admits a Novikov structure by Proposition
$\ref{iso}$. Hence we may assume $\ga_{i,0}=0$ for all $i$.
But then we may also assume that $\ga_{i,1}=0$ for $i=2,\ldots ,m$.
Just consider the new basis $(e_1,e_2-\ga_{2,1}e_1,\ldots ,e_m-\ga_{m,1}e_1)$
for $\Lb$. 
Now we claim that
\eqref{produkt} defines a Novikov product, where $\phi_1=0,
\, \phi_2=\phi$, the product on $\La$ and $\Lb$ is trivial, 
$A_1=\phi (e_1)=J(n)$ and
\begin{align*}
X_i & = 0,\; i=1,\ldots ,m\\
x_{11} & =0 \\
x_{1j} & = v_{1j},\; j=2,\ldots ,m\\
x_{ij} & = A_1^tA_iv_{1j},\; i=2,\ldots ,m,\; i\le j\\
x_{ji} & = x_{ij}-v_{ij},\; j>i 
\end{align*}
We first show that
\begin{align}
A_1A_1^tA_j & = A_j \label{33} \\
A_iA_1^tA_j & = A_jA_1^tA_i \label{34}
\end{align} 

for all $i,j$: we have $A_1A_1^tA_1=A_1$ since $A_1=J(n)$. It follows
$A_1A_1^tA_1^k=A_1^k$ for all $k\ge 1$. Since each $A_j$ is a polynomial
in $A_1$ the first claim follows. Similarly $A_iA_1^tA_1=A_i$ so that
$$
A_iA_1^tA_1^k=A_iA_1^{k-1}=A_1^{k-1}A_i=A_1^kA_1^tA_i
$$
Since each $A_j$ is a polynomial in $A_1$ the second claim follows.
Recall that we assume \eqref{24}. Since $X_i=0$ for all $i$ we only have
to show that \eqref{27} is satisfied:
\begin{align*}
A_ix_{jk} & = A_j x_{ik}
\end{align*} 
for all $i,j,k$ in $[1,m]$. According to the definition of $x_{ij}$
we have to consider different cases.
Let us first assume that $i,j,k\ge 2$.  
The first case is $i\le k, j\le k$. 
We have to show
$$
A_iA_1^tA_jv_{1k} = A_jA_1^tA_iv_{1k}
$$
which follows from \eqref{34}.
The second case is $i\le k, j>k$. We have to show
$$
A_i(A_1^tA_kv_{1j}-v_{kj}) = A_jA_1^tA_iv_{1k}
$$
Using \eqref{34} and \eqref{24} we obtain
\begin{align*}
0 & = (A_iA_1^t)(A_kv_{1j}+A_1v_{jk}-A_jv_{1k}) \\
 & = A_iA_1^tA_kv_{1j}-A_iA_1^tA_1v_{kj}-A_iA_1^tA_jv_{1k}\\
 & = A_iA_1^tA_kv_{1j}-A_1A_1^tA_iv_{kj}-A_jA_1^tA_iv_{1k}\\
 & = A_i(A_1^tA_kv_{1j}-v_{kj})-A_jA_1^tA_iv_{1k}
\end{align*}
The third case is $k<i, j>k$. Then we have to show
$$
A_i(A_1^tA_kv_{1j}-v_{kj}) = A_j(A_1^tA_kv_{1i}-v_{ki})
$$
We have
\begin{align*}
0 & = (A_jA_1^t)(A_iv_{1k}-A_kv_{1i}+A_1v_{ki}) \\
 & = A_jA_1^tA_iv_{1k}-A_jA_1^tA_kv_{1i}+A_jA_1^tA_1v_{ki}\\
 & = A_iA_1^t(A_jv_{1k})-A_jA_1^tA_kv_{1i}+A_jv_{ki}\\
 & = A_iA_1^t(A_kv_{1j}-A_1v_{kj})-A_jA_1^tA_kv_{1i}+A_jv_{ki}\\
 & = A_i(A_1^tA_kv_{1j}-v_{kj}) - A_j(A_1^tA_kv_{1i}-v_{ki})
\end{align*}
Finally suppose that one of the indices is equal to one.
Assume that $i=1$. If $j\le k$ we have to show
$$
A_1(A_1^tA_jv_{1k}) = A_jv_{1k}
$$
which follows from \eqref{33}. If $j>k \ge 2$ we have to show
$$
A_1(A_1^tA_kv_{1j}-v_{kj}) = A_jv_{1k}
$$
which follows from \eqref{24}. For $k=i=1$ there is nothing to show.
Similarly the other cases are shown.
\end{proof}

\begin{cor}
Any filiform Lie algebra with abelian commutator algebra admits a
Novikov structure.
\end{cor}

\section{A reduction to the case of nilpotent Lie algebras}
In the previous section, we established some examples of LSA and 
Novikov structures on several classes of Lie algebras. 
All of these examples were nilpotent. One 
might think that these cases are too special and that we should look for  
solvable, but non-nilpotent Lie algebras. However, in this section we will 
show that the lifting procedure can in many cases be reduced to the nilpotent case.

To obtain this result, we need some preliminary work on the structure of 
modules over a nilpotent Lie algebra.
\begin{lem}\label{column-row}
Let $\Lg$ be a finite dimensional nilpotent Lie algebra over a field $k$
of characteristic 0, equipped with a representation $\varphi:\Lg\rightarrow
\Lg\Ll(n,k)$. Then $k^{n\times 1}$ (resp.\ $k^{1\times n}$) becomes a
$\Lg$-module, via $X\cdot v = \varphi(X) v,\; \forall X\in \Lg, \forall v
\in k^{n\times 1}$ (resp.\ $X\cdot v =-v \varphi(X),\; \forall X\in \Lg,
\forall v \in k^{1\times n}$). We then have that, for these module
structures,
\[ H^0(\Lg,k^{n\times 1})=0 \Longleftrightarrow H^0(\Lg,k^{1\times n})=0\]
\end{lem}

\begin{proof} Recall that  $H^0(\Lg,M)$ is
the space $M^\Lg=\{m\in M\; | \; X\cdot m=0,\;\forall X\in \Lg\}$.\\
Suppose first that $H^0(\Lg,k^{n\times 1})=0$ and assume that there exists
a non-zero $w\in k^{1\times n}$ for which $- w \varphi(X)=0$. Then, after
conjugating with an invertible $n\times n$-matrix, we may suppose that
\[\varphi(X)= \left(\begin{array}{c|cccc}
0 & 0 & 0 & \cdots & 0 \\ \hline
* &   &   &        &   \\
* &   &   &        &   \\
\vdots  &   &   &  *      &   \\
* &   &   &        &
\end{array}\right),\;\forall X \in \Lg.\]
Now, let $V\subseteq k^{n\times 1}$ be the subspace consisting of all
column-vectors $v=(0,a_2,a_3,\ldots,a_n)^t$ having 0 as their first
coordinate. It is obvious that $V$ is a $\Lg$-submodule of $k^{n\times 1}$
and there is a short exact sequence of $\Lg$-modules
\[0 \rightarrow V \rightarrow k^{n\times 1}\rightarrow k \rightarrow 0\]
where $k$ stands for the trivial 1-dimensional $\Lg$-module. This short
exact sequence gives rise to a long exact sequence in cohomology:
\[0\rightarrow H^0(\Lg, V) \rightarrow H^0(\Lg,k^{n\times 1})\rightarrow
H^0(\Lg,k) \rightarrow H^1(\Lg,V)\rightarrow \cdots\] As
$H^0(\Lg,k^{n\times 1})=0$, we also have that $H^0(\Lg,V)=0$. Because
$\Lg$ is a finite-dimensional nilpotent Lie algebra, this implies that
$H^i(\Lg,V)=0$, for all $i\geq 0$ (see \cite{DEK}). It follows that in the
exact sequence above, $H^0(\Lg,k)$ is standing between two $0$-terms and
hence must itself be 0. But this contradicts the fact that for the trivial
module, we have that $H^0(\Lg,k)=k$.\\
The other direction can be proved in a similar way.
\end{proof}

Let $\Lg$ be a Lie algebra over a field $k$ and assume that
$\varphi_1:\Lg\rightarrow\Lg\Ll(n_1,k)$ and $\varphi_2:\Lg\rightarrow
\Lg\Ll(n_2,k)$ are two Lie algebra representations. Then $k^{n_1\times
n_2}$ becomes a $\Lg$--module via the map
\[ \varphi: \Lg \rightarrow \End(k^{n_1\times n_2}): X \mapsto \varphi(X)\]
where
\[\forall B \in k^{n_1\times n_2}:\; \varphi(X)(B) = \varphi_1(X) B - B \varphi_2(X). \]
We refer to $\varphi$ as being the {\it combination} of $\varphi_1$ and
$\varphi_2$.
\begin{lem}\label{combination}
Let $\Lg$ be a finite dimensional nilpotent Lie algebra over a field $k$
of characteristic $0$. Assume that $\varphi_1:\Lg\rightarrow\Lg\Ll(n_1,k)$
and $\varphi_2:\Lg\rightarrow \Lg\Ll(n_2,k)$ are two Lie algebra
representations, such that
\begin{enumerate}
\item $\varphi_1(\Lg)$ consists of nilpotent matrices and
\item $H^0_{\varphi_2}(\Lg, k^{n_2\times 1}) =0.$
\end{enumerate}
If $\varphi$ denotes the combination of $\varphi_1$ and $\varphi_2$, then
\[H^i_\varphi(\Lg,k^{n_1\times n_2})=0,\;\forall i\geq 0.\]
\end{lem}
\begin{proof}
Note that by \cite{DEK} again, we only need to prove the lemma for $i=0$.\\
We proceed by induction on $n_1$. If $n_1=1$, then $\varphi_1\equiv 0$ and
the claim follows from lemma~\ref{column-row}.\\
Now, let $n_1>1$. Without loss of generality, we may assume that $\forall
X\in \Lg:\;  \varphi_1(X)$ is an upper-triangular matrix with 0's on the
diagonal. Let $B=(b_{i,j})\in k^{n_1\times n_2}$ be an element with
\[0=\varphi(X) B= \varphi_1(X)B-B\varphi_2(X),\;\forall X\in \Lg.\]
Looking at the last row of the above equality, we find
\[ (0,0,\ldots,0)=(0,0,\ldots,0) - (b_{n_1,1}, b_{n_1,2},\ldots,b_{n_1,n_2})\varphi_2(X), \;\forall X\in
\Lg\] From the case $n_1=1$ (or the previous lemma), we obtain that the
last row of $B$ has to be the zero row. The rest now follows by induction.
\end{proof}

\begin{prop}\label{sum-mod}
Let $\Lg$ be a nilpotent Lie algebra over a field of characteristic
0 and let $V$ be
a finite dimensional $\Lg$-module. Then $V$ can be written as a direct sum
\[V_n\oplus V_0,\]
where $V_n$ is a nilpotent $\Lg$-module (in fact the unique maximal
nilpotent submodule of $V$) and $V_0$ is also a $\Lg$-module with
$H^i(\Lg,V_0)=0$, $\forall i\geq 0$.
\end{prop}
\begin{proof}
It is easy to see that $V$ contains a unique maximal nilpotent submodule
$V_n$. Therefore, after choosing a suitable basis for $V$, we can assume
that the module structure is given in matrix form $\psi:\Lg \rightarrow
\Lg\Ll(n_1+n_2,k)$ ($n_1+n_2$ is the dimension of $V$), with
\[\forall X\in \Lg:\; \psi(X)= \left(
\begin{array}{cc}
\varphi_1(X) & B_X \\
0 & \varphi_2(X)
\end{array} \right) \]
where $\varphi_1:\Lg\rightarrow \Lg\Ll(n_1,k)$ has images inside the set
of upper triangular matrices with 0's on the diagonal and
$\varphi_2:\Lg\rightarrow \Lg\Ll(n_2,k)$ is such that
$H^0_{\varphi_2}(\Lg,k^{n_2\times1})=0$
(and hence $H^i_{\varphi_2}(\Lg,k^{n_2\times1})=0$ as before).\\
Let $\varphi$ denote the combination of $\varphi_1$ and $\varphi_2$ as
above. We claim that the map $B:\Lg\rightarrow k^{n_1\times n_2}: X
\mapsto B_X$ is a 1-cocycle (with respect to the module structure
determined by $\varphi$). To see this, we must check that
\begin{eqnarray*}
B_{[X,Y]} & = & \varphi(X) B_Y - \varphi(Y) B_X\\
& = & \varphi_1(X) B_Y - B_Y \varphi_2(X) - \varphi_1(Y) B_X + B_X
\varphi_2(Y)
\end{eqnarray*}
This is easily seen to hold, by writing out the identity
$\psi([X,Y])=\psi(X)\psi(Y)-\psi(Y)\psi(X)$. By lemma~\ref{combination},
we know that $H^1_\varphi(\Lg,k^{n_1\times n_2})=0$ and hence $B$ has to a
coboundary. Therefore, there exists a $\alpha\in k^{n_1\times n_2}$ such
that
\[ B_X=\varphi(X)\alpha= \varphi_1(X)\alpha -\alpha\varphi_2(X).\]
If we now conjugate $\psi$ with the matrix $\left(\begin{array}{cc}
I_{n_1} & \alpha \\ 0 & I_{n_2}
\end{array}\right)$, we get
\begin{eqnarray*}
\left(\begin{array}{cc} I_{n_1} & \alpha \\ 0 & I_{n_2}
\end{array}\right) \psi(X)\left(\begin{array}{cc}
I_{n_1} & -\alpha \\ 0 & I_{n_2}
\end{array}\right) & = & \left(\begin{array}{cc}
I_{n_1} & \alpha \\ 0 & I_{n_2}
\end{array}\right)\left(\begin{array}{cc}
\varphi_1(X) & B_X \\ 0 & \varphi_2(X)
\end{array}\right)\left(\begin{array}{cc}
I_{n_1} & -\alpha \\ 0 & I_{n_2}
\end{array}\right)\\
&=&\left(\begin{array}{cc} \varphi_1(X) & 0 \\ 0 & \varphi_2(X)
\end{array}\right)
\end{eqnarray*}
This shows that we can decompose $V$ into two submodules as claimed.
\end{proof}

We are now ready to apply our results to the lifting problem of LSA or
Novikov structures.
Let $\La$ be an abelian Lie algebra as before and assume that $\Lb$ is 
a nilpotent Lie algebra. Let $\varphi:\Lb\ra \End(\La)$ denote a $\Lb$-module 
structure of $\La$. By Proposition~\ref{sum-mod}, we can write $\La$ as 
a direct sum of $\Lb$-modules $\La=\La_n\oplus \La_0$, where $\La_n$ is a 
nilpotent $\Lb$-module and $H^0(\Lb,\La_0)=0$. We can write 
$\varphi=\varphi'\oplus\varphi''$, where $\varphi'$ (resp.\ $\varphi''$) is the 
restriction of $\varphi$ to $\La_n$ (resp.\ $\La_0$). It follows that for any
$i$, we have that 
\[H^i(\Lb,\La)=H^i(\Lb,\La_n)\oplus H^i(\Lb,\La_0)=H^i(\Lb,\La_n)\oplus 0=
H^i(\Lb,\La_n).\]
Any 2-cocycle $\Omega\in Z^2(\Lb,\La)$ is a sum of two 2-cocycles
\[\Omega=\Omega'+\Omega'',\mbox{ with }
\Omega'\in Z^2(\Lb,\La_n)\mbox{ and }\Omega''\in Z^2(\Lb,\La_0).\]
As $H^2(\Lb,\La_0)=0$, we have that $\Omega$ is in fact 
cohomologuous to $\Omega'$. Now, let $\Lg$ denote the extension of $\Lb$ 
by $\La$ determined by the 2-cocycle $\Omega$. Without loss of generality
we may assume that $\Omega''=0$. Note that $\Omega'$ determines an extension
$\Lg_n$ of $\Lb$ by $\La_n$ which is in fact given by 
\begin{equation}\label{nilpotent}
0\rightarrow \La_n=\La/\La_0\rightarrow \Lg_n=\Lg/\La_0\rightarrow 
\Lb\rightarrow 0
\end{equation}
As $\Lb$ is nilpotent and $\La_n$ is a nilpotent $\Lb$-module, we have that 
$\Lg_n$ is a nilpotent Lie algebra. We will refer to the extension 
(\ref{nilpotent}) as the nilpotent extension induced by the extension 
$\Lg$. We can now formulate our reduction statement.
\begin{prop}\label{reduction}
Let $\La$ be an abelian Lie algebra equipped with a trivial LSA-product and 
let $\Lb$ be a nilpotent Lie algebra equipped with a left-symmetric structure.
Let $\Lg$ denote an extension of $\Lb$ by $\La$ and let $\Lg_n$ be the 
induced nilpotent extension. If the left symmetric-structure of $\Lb$ 
lifts to the extension $\Lg_n$, then it also lifts to $\Lg$.
\end{prop}
\begin{proof}
As above, we can assume that $\Lg$ is determined by a cocycle $\Omega=
\Omega'$, with $\Omega(x,y)\in \La_n$, $\forall x,y\in \Lb$. 
The extension $\Lg_n$ is represented by the same 2-cocyle, but then seen as
an element of $Z^2(\Lb,\La_n)$.\\
We have to show that the conditions of Proposition~\ref{phi12} are satisfied 
for $\Lg$, given they are satisfied for $\Lg_n$. Let 
\[\omega':\Lb\times \Lb\rightarrow \La_n\mbox{ and }
\varphi_1',\varphi_2':\Lb\rightarrow \End(\La_n)\]
denote the maps needed to fulfil all the condition of Proposition~\ref{phi12} 
for the extension $\Lg_n$. Now, let 
\[ \omega'':\Lb\times \Lb\rightarrow \La_0:(x,y)\mapsto 0\mbox{ and }
\varphi_1'',\varphi_2'':\Lb\rightarrow \End(\La), \mbox{ with }
\varphi_2''=\varphi''\mbox{ and } \varphi_1\equiv 0.\]
It is now obvious that 
\[ \omega=\omega'+\omega'',\;\varphi_1=\varphi_1'+\varphi_1''\mbox{ and }
\varphi_2=\varphi_2'+\varphi_2''\]
satisfy all of the conditions of Proposition~\ref{phi12} for the extension
$\Lg$.
\end{proof}
\begin{rem}
This proposition provides an alternative proof for Proposition~\ref{iso},
because the conditions of this proposition immediately imply that 
$\La_n=0$, hence $\Lg_n=\Lb$.
\end{rem}
Analoguously, we find
\begin{prop}
Let $\La$ and $\Lb$ be abelian Lie algebras equipped with a trivial 
LSA-product and
let $\Lg$ denote an extension of $\Lb$ by $\La$ with induced nilpotent 
extension $\Lg_n$. If we can lift the trivial LSA-structures to a 
Novikov structure on $\Lg_n$, then we can also lift it to 
a Novikov structure on $\Lg$.
\end{prop}
\begin{proof}
The proof is analoguous to the previous proposition, but now we have to
satisfy the conditions of Corollary~\ref{trivial}.
\end{proof}
As an application of the reduction we obtained in this section, we prove 
the following generalization of Scheuneman's result.
\begin{prop}\label{prop5.7}
Let $\Lg$ be a 2-step solvable Lie algebra with 
$\Lg^r=\Lg^4$ for all $r\ge 5$. Then $\Lg$ admits a complete left symmetric structure.
\end{prop}
\begin{proof}
$\Lg$ can be seen as an extension of abelian Lie algebras
\[ 0 \rightarrow [\Lg, \Lg] \rightarrow \Lg \rightarrow \Lg/[\Lg,\Lg]
\rightarrow 0.\]
As a $\Lg/[\Lg,\Lg]$-module, we have a decomposition 
$[\Lg,\Lg]=[\Lg,\Lg]_n\oplus[\Lg,\Lg]_0$ and the resulting quotient 
$\Lg_n=\Lg/[\Lg,\Lg]_0$ is a nilpotent Lie algebra. Hence there exists a 
$c$ such that $\Lg^c\subseteq [\Lg,\Lg]_0 $ and thus certainly 
$\Lg^4\subseteq [\Lg,\Lg]_0$. It follows that $\Lg_n$ is nilpotent of class
$\leq 3$. By Proposition~\ref{scheuneman} and its proof, 
we know that the trivial 
LSA-structures on $\La_n=[\Lg,\Lg]_n$ and $\Lb=\Lg/[\Lg,\Lg]$ lift to 
an LSA structure on the extension $0\ra \La_n \ra \Lg_n \ra \Lb \ra 0$. By 
Proposition~\ref{reduction}, we can conclude that we also obtain an 
LSA-structure on $\Lg$.
\end{proof}

\end{document}